# Quantum dynamics of molecules in $^4$He nano-droplets: Microscopic Superfluidity

S Dey[*,1,2], J P Gewali[2], A K Jha[2], L Chhangte[2] and Y S Jain[2]

[1]Don-Bosco College of Engineering and technology, Assam Don-Bosco University,
Guwahati-781 017, Assam, India

[2]Department of Physics, North-Eastern Hill University,
Shillong-794 022, Meghalaya, India

E-mail : samratthegr8@gmail.com

**Abstract** : High resolution spectroscopy of doped molecules in $^4$He nano-droplets and clusters gives a signature of superfluidity in microscopic system, termed as *microscopic superfluidity*. Ro-vibrational spectrum of $^4$He$_N$-M clusters is studied with the help of some important observations, revealed from experiments (*viz*., localised and orderly arrangement of $^4$He atoms, although, being free to move in the order of their locations; individual $^4$He atoms can not be tagged as normal/ superfluid, *etc.*) and other factors (e.g., consideration that the $^4$He atoms which happen to fall in the plane of rotation of a molecule, render a equipotential ring and thus, do not take part in rotation; *etc.*) which effect the rotational and vibrational spectrum of the system. This helps us in successfully explaining the experimental findings which state that the rotational spectrum of clusters have sharp peaks (indicating that the molecule rotates like a free rotor) and moment of inertia and vibrational frequency shift have a non-trivial dependence on *N*.

## 1. Introduction

Superfluids (characterised by unusual properties, like zero viscosity, *etc.*) represent the effects of quantum mechanics acting on the bulk properties of matter on a large scale and thus, superfluidity is, rightly, termed as *macroscopic quantum phenomenon* [1]. Recently, however, the high resolution spectroscopy of doped molecules in $^4$He nano-droplets (number of $^4$He atoms, N ~ $10^3 - 10^4$) and clusters ($^4$He$_N$–*M*, M being the embedded molecule and *N* = 1, 2, ...) [2] provided a unique method to explore superfluidity in such a microscopic system (*MICROSCOPIC SUPERFLUIDITY* [3]), as well [4–6]. It is found that the rotational spectrum of an embedded molecule in liquid $^4$He shows sharp peaks, similar to that of the molecule in gaseous state [5, 7, 8] which indicates that the molecule rotates like a free rotor, although, with an increased moment of inertia ($I$). In addition, one also observes a non trivial dependence of $B$ (rotational constant) and vibrational frequency shift on $N$ whose details depand on the constitution of the embedded molecules [2, 9] (with $N_2O$ and HCCCN as dopent molecules, respectively). Consequently, numerous studies (both theoretical and experimental) on such high resolution spectroscopy have been reported till date [2–32]. Such studies are considered very significant from the point of view of tracing the evaluation of superfluidity, through an intermediate range between the limiting cases of an individual molecule at one end and bulk matter at the other [2]. However, no theoretical approach could give a sound explanation of the experimental findings, till date. So, we analyse the spectrum of an extensively studied linear molecule, $N_2O$ (structure shown in Fig. 1(a), for which the turn around of $B$ with $N$ was observed for the first time, with the help of some simple but concrete physical arguments. We, further, verify our approach by considering another molecule, HCCCN (structure shown in Fig 5(a).

---
[*]Corresponding Author

## 2. Experimental studies

Chronologically, the first spectroscopic experiment in which different atoms were implanted into the interior of liquid $^4$He was reported in 1985 [33, 34]. The first similar study of a molecule ($SF_6$) with low resolution infrared spectrum was reported in 1992 [35], followed by its high resolution spectroscopic study in 1995 [4]; the term *high resolution* spectra refers to rotationally resolved microwave/ infrared spectra. In 1998, Grebenev *et al*. analysed the spectra of a molecule, embedded in as few as 60 $^4$He atoms [5]. However, it was not before 2002 that the first $N$ dependence of vibrational and rotational spectra (B) was determined (with OCS as the probe molecule) using IR diode laser spectrometer and molecular beam Fourier-transform microwave spectrometer for the analysis of infrared and microwave spectra, respectively [29]; in such a case, the cluster was generated by supersonic jet expansions (from a cooled nozzle) of trace amount of the molecule of interest in $^4$He with high backing pressure [30]; by tuning the pressure and temperature in the production chamber, the relative abundance of a given cluster (of a particular N) is controlled, allowing for the assignment of individual lines to such clusters; the method is remarkably different from the earlier techniques, producing cluster/ nano-droplets of large N, *e.g.*, the Helium Nano-droplet Isolation (HENDI) spectroscopy [7] (where $^4$He nano-droplets, typically consisting of several thousand $^4$He atoms, were produced first and then made to pick up the molecule of interest), microscopic Andronikashvili experiment [5] (where they could make a solvation of OCS by about 60 $^4$He in a $^3$He droplet), *etc.* The turn around of $B$ with $N$ was observed for the first time in 2003, with $N_2O$ as the dopent molecule. Since then, along with OCS [14] and $N_2O$ [2, 3, 15–20], the vibrational and rotational spectra of several other molecules, *e.g.*, HCCCN [9], $CO_2$[31], CO[32], *etc.*, for very small N, have also been reported.

Although, the reports [3, 15–18] give the earlier high resolution spectroscopic data for our molecule of interest, $N_2O$, in $^4$He, our present analysis uses very recent experimental results of Mckellar, published in 2007 [2]. $B$ decreases monotonically with $N$ increasing from 1 to 6. While, it remains nearly constant for $N$ = 6 to 8, it increases and decreases in a periodic manner for further increase in $N$. Similarly, vibrational frequency exhibits blue shifts for $N$ = 1 to 5 and a red shift thereafter, with a change in slope at $N \sim 17$. It is needless to mention that each rotational spectrum has sharp peaks, similar to that of the molecule in gaseous state, indicating that the molecule rotates like a free rotor, although, with an decreased $B$. While verifying our approach with another molecule (HCCCN), we have again used another recent paper [9]; although, unlike $^4He_N$-$N_2O$ clusters, we do not have data for vibrational frequency for different $N$ in case of $^4He_N$-HCCCN clusters, we find that our modus operandi to explain the rotational dynamics of HCCCN is similar to that of $N_2O$. Rotational transition frequency decreases, almost monotonically, till $N$ = 9 remains nearly constant for $N$ = 9 to 11 and then, increases and decreases, giving an indication of oscillatory nature, like that for $N_2O$ molecule. However, for $N = 18 - 25$, $J = 1 - 0$ transitions were of too low in frequency to be measured and $J = 2 - 1$ transitions were of, apparently, very low intensity.

### 3. Theoretical studies

Various phenomenological models and computer based simulation techniques have been applied to explain the free rotation (with an increase in $I$) of an embedded molecule in nano-droplets, earlier [4, 5, 7, 10, 27, 36, 37] and the $N$ dependence of vibrational frequency and $B$ of such a molecule in clusters, thereafter [6, 21–26]. While, the theoretical models were predominantly used to explain the free rotation, with reduced $B$, of a molecule in $^4$He nano-droplets, recently, the simulation techniques have also become successful, apparently, to explain the N dependence of vibrational frequency and $B$ of such a molecule in clusters, to a certain extent.

*3.1. Models and their limitations :*

Among the models, the one known as supermolecule (or, donut) model [4], assumes that a specific number of $^4$He atoms rotate rigidly with the molecule leading to an increase in $I$. However, this model suffers from two problems. (1) In case of lighter molecules, such as, $NH_3$ and HCN, even a single rigidly attached $^4$He atom increases $I$ by an order of magnitude more than what is observed experimentally [7]. (2) The results of OCS-$^4H_2$ complex in $^4$He are inconsistent with the donut model; it was found by isotopic substitution that a single $H_2$, HD or $D_2$ molecule will replace one $^4$He in the donut and extrapolation to zero mass of the isotopomer gives a $I$ of the donut with a portion of it taken out considerably larger than for the OCS in a pure $^4$He droplet, *i.e.*, with a complete donut [7].

A more sophisticated *two fluid model* [5], takes note of Landaus's two fluid phenomenology and relates the observation of free rotation of the embedded molecule with superfluidity of $^4$He. It assumes that the density of $^4$He atoms around the molecule can be partitioned into spatially dependent normal fluid and superfluid fraction; the former, rotating rigidly with the molecule, has a large value in the first solvation layer. Thus only the normal fluid fraction contributes to *I*. However, this model was incomplete, as the authors provided no definition of the spatially dependent normal fluid density [7]. Thus, a slightly different version of two fluid model was introduced [36] which used the term *nonsuperfluid density* (= the difference in total density and superfluid density) instead of normal *fluid density*, because, the density is not related to the normal fluid density produced by a gas of elementary excitations, as in bulk $^4$He experiments, such as, the Andronikashvili experiment [7, 28]. Yet another model, known as *quantum hydrodynamic model* [7, 10, 27], relies on two assumptions, *viz*., (i) $^4$He is fully superfluid, and (ii) $^4$He density is supposed to rearrange instantaneously around the rotating molecule by an irrotational flow. In others words, the density in the rotating frame of the molecule is constant and equal to that of the static case; the density profile is derived from density functional calculations. The solvent rearrangement is defined as adiabatic following. In this context, it may be noted that a linear motion of a rigid body through an ideal (aviscous and irrotational flow) fluid generates motion in the fluid that contributes to the effective mass for translation. Similarly, rotation of an ellipse generates $^4$He kinetic energy, proportional to the square of the angular velocity, thus contributing to the effective *I*. While, this factor has already been incorporated in [5], attempts to estimate this term, using classical expressions for *I* of an ellipsoid in a uniform fluid of the density of bulk $^4$He, gave results which were small fraction of the experimental value. However, quantum hydrodynamic model properly takes into account the anisotropy of the density and has obtained *I* in good agreement with experiments for a number of molecules. But, as stated in [28], for light rotors, for which the hypothesis of adiabatic following does not hold, the model overestimates the *I* increase; experimentally light rotors are found to have no change in *I* [12].

Apart from the limitations of these models stated above, one may also note that all these models were, indeed, proposed much before 2002, when the first *N* dependence of the spectra of a dopent molecule in a cluster was determined [29]. Consequently, the said *N* dependence was not taken care of in these theoretical models.

*3.2 Simulations and Their Limitations :*

It may be noted that various computer simulation techniques were also applied earlier to investigate the free rotor behaviour (with an increased *I*) of an embedded molecule [37], including some of the techniques used in the theoretical models [*e.g*., Path Integral Monte Carlo (PIMC) was applied in two fluid model by [36]; in fact, simulation of pure $^4$He began as early as 1983 [38] and the first such calculation of molecule doped in $^4$He was reported in 1992 [39]. However, recent progress in Quantum Monte Carlo (QMC) simulations are now allowing a direct access to the dynamics of molecules in clusters with very few $^4$He atoms [28] and consequently, providing a better understanding of *N* dependence of vibrational and rotational spectra (*B*).

Generally, two variants of QMC, Diffusion Monte Carlo (DMC) and PIMC, have been being used to study a dopent molecule [37]. DMC again possesses two variants, *viz*., fixed node DMC and Projection Operator Imaginary Time Spectral Evolution (POITSE) [37]. The pioneering study in the direction of the said *N dependence* was done in 2003 [21, 22], soon after their experimental discovery in 2002 [29]. While in [21] POITSE was applied, in [22] Reptation Quantum Monte Carlo (RQMC) technique, a combination of some of the features of PIMC and DMC, was used. In [22] it was identified that tunnelling is the key to understand the onset of superfluidity. It was suggested that as soon as the doughnut ring occupation is completed, both polar regions start to be populated. Once the closure of $^4$He rings in the *sagittal* planes is over, atomic exchanges along these rings become possible, without contributing to *I*. These exchanges become possible as result of the tunnelling through the potential barrier which separates the *equatorial* region from the two *polar* regions; larger the barrier, the smaller the tunnelling and larger the corresponding effective *I* will be. This approach was further developed and discussed in [23] in 2005. In [21], a partial adiabatic following of $^4$He is considered and the turn around of *B* was related to the onset of permutation exchanges between $^4$He atoms. This method is also refined and described in some recent papers, *e.g.*, [24]. In 2004, Blinov *et al.* used PIMC, incorporating the rotational degrees of freedom of the dopent molecule, to study the dynamics of doped clusters [25] and explained the decrease of *I* with increase in *N*, with an argument similar to [22]. Recently in 2005, Zillich *et al.* [26] have developed a PIMC method to treat rotating molecules in superfluids. Independently, Miura developed a Path Integral Hybrid Monte Carlo (PIHMC), to handle the rotational motion of the dopent molecules quantum mechanically, in the same year [40].

Thus, simulations tried to solve the problem of analysis of the *N* dependence of vibrational and rotational spectra of different molecules. However, not only the simulation techniques are computationally expensive and are, thus, limited to few $^4$He atoms, but also the results show significant variation among themselves (different techniques) as well as with experiments [6, 14, 21–24, 26]. The simulations could not reproduce the the oscillatory behaviour in the *N* dependence of *B* [14]. Also, the simulation techniques are very sensitive to the quality of $^4$He-M potential [23] and different forms of potentials were tried to reproduce the experimental results, but in vain. Furthermore, different arguments used in the simulation techniques are not acceptable from the physics point of view. For example, while using RQMC in [22], it was stated that *The number of He atoms near one end is 0.2, 0.3, and 0.7, for n = 6, 7, and 8, and 0.5, 0.8, and 1.1 near the other end* which is, indeed, against the principle of indivisibility of atoms. Similarly, in [6, 21] the presence of two fluids (*superfluid and nonsuperfluid*) were mentioned to exist even in clusters as small as that of containing only one solvation shell which is, again, not acceptable from physical point of view.

For $N_2O$ embedded in $^4$He clusters, simulation by Moroni *et al.* used RQMC [19], while, that of Paseani *et al.* used POITSE [20]. While, the former deviated largely from the experimental trend in the large *N* region *i.e.*, for *N* = 25 and 30 (although, it had a little success for small *N*), the later failed even in the small *N* region, especially for *N* = 10 − 16. For HCCCN embedded in $^4$He clusters, the results of the simulation vary significantly with those of the experiments; when the form of interaction potential is that of [41], the simulation result is worse in case of larger clusters and when the form of interaction potential is that of [42], the simulation result is worse in case of smaller clusters. This, together with what is discussed in the last paragraph, clearly reveals the limitations.

**4. Need for an alternative approach**

In what follows from the above discussion, we indeed lack a viable approach that comprehensively explains different aspects of experimental observations of a dopent molecule in cluster/ nano-droplet; this is what is also stated in [7, 8, 14]. Consequently, we make some important observations. We note that $^4$He atoms around the embedded molecule are localised in very tiny space (of the order of intermolecular distances) where it is impossible to imagine any $p = 0$ condensate and where the number of $^4$He atoms are so few that considering them to be a mixture of superfluid and normal fluid $^4$He atoms will mean tagging each atom as being superfluid or normal fluid which is absurd, as the $^4$He atoms are identical. Furthermore, we note that: (i) the small magnitude of total blue shift (which gets even smaller with increase in $N$) ([2]) indicates that $^4$He-M interaction is very weak. (ii) The nature of the spectroscopic ro-vibrational structure of the embedded molecule is similar to that of the molecule in gas phase [7, 8] which indicates that $^4$He atoms around the molecule, to a good approximation, render isotropic potential for rotation of the rotor. So, we consider that the $^4$He atoms are having an ordered arrangement. (iii) The sharp rotational and vibrational spectral lines [5, 7, 8] indicate the fact that $^4$He atoms in $^4$He$_N$-M do not have relative motion with respect to the rotor (the part of the cluster which is rotating) or the vibrating $M$. Consequently, we consider that the $^4$He atoms are localised (with zero point uncertainty in position) and form a structure which does not change with time, as only this ensures that the rotor or the vibrating molecule see a time independent potential, necessary for sharp spectral lines; to a good approximation, this requirement is satisfied even if the particles around the rotor move in the order of their locations. In short, we consider a localised and orderly arrangement of identical $^4$He atoms to explain this microscopic superfluidity, in detail.

**5. Analysis**

As mentioned above, we, basically, consider a localised and orderly arrangement of $^4$He atoms to explain the experimental findings. Further, we also introduce the idea of *equipotential ring* in the sagittal plane of the molecule, because, although it is well known that each $^4$He atom added to a cluster, $^4$He$_N$-M, can render new bond lengths and bond angles providing a different structure and symmetry of $^4$He$_{N+1}$-M which obviously has different $I$, it is seen that this simple fact is not sufficient to account for experimentally observed decrease in $I$ with increasing $N$. We consider that those $^4$He atoms in the cluster which happen to be in the molecular plane render a sort of equipotential ring for a change in the angular posture of therest of the cluster which can, therefore, rotate like a free rotor. This consideration finds strong support from the fact that $^4$He atoms in different rings in a quantum vortex move with different speeds without interfering with each other [43]. Thus, each extra $^4$He atom, joining other atoms in the said equipotential ring, does not increase $I$. On the contrary, as the number of $^4$He atoms increases, the order of the rotational symmetry ($n$) of the potential, $V_n = V_0/2[1 - \cos(n\phi)]$, also increases which tends to decrease $I$, as evident from Mathieu's equation [44] (the point is elaborated in the next subsection); we assume $V_0 \ll 2B$, in conformity with the observed free rotation. In fact, Mathieu's equation also suggests that the increase in the height of the potential hills, $V_0(\ll 2B)$, also decreases $I$. We use these inferences to explain the experimental observations of $^4$He$_N$-N$_2$O, first and verify the approach by $^4$He$_N$-HCCCN, later.

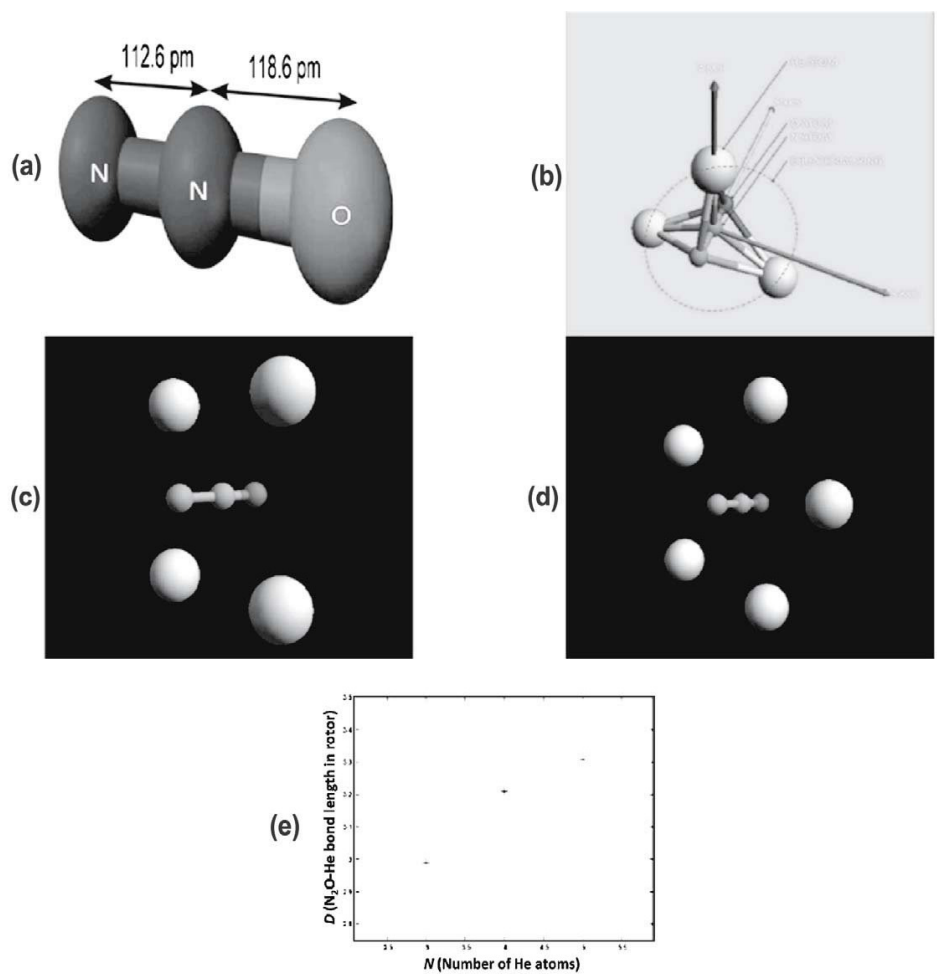

**Figure 1.** (a) The structure of N$_2$O molecule. (b) 3 $^4$He atoms [in the equatorial plane (z-x plane)] attached to N$_2$O and rotating with it along an axis perpendicular to y axis. (c) 4 $^4$He atoms (in the equatorial plane) attached to N$_2$O and rotating with it. (d) 5 $^4$He atoms (in the equatorial plane) attached to N$_2$O and rotating with it. (e) Change of D (in Å) with $N$ for $N = 3 - 5$.

*5.1. Analysis of $^4He_N$-$N_2O$ :*

Identifying the $N_2O$ molecular axis with the y axis and that the molecule rotates about z axis, $^4$He atoms are first considered to occupy positions on a ring (called as the equatorial ring) in x-z plane. These atoms are assumed to have reasonably good binding with the molecule and consequently, the resulting complex rotates like a rigid rotor (as shown in Fig. 1(b), 1(c) and 1(d) for $N$ = 3, 4 and 5, respectively). Obviously, $I$ of the rotor increases ($B$ decreases) with increase in $N$ of such atoms, in consistence with [2]. Furthermore, the force constant of the vibrating molecule can also be assumed to increase with $^4$He atoms coming in the equatorial ring, as they are expected to strengthen the binding between the atoms of the molecule (*i.e.*, the relative position between two atoms), if one considers that the $^4$He atoms, in this position, are bound to all the atoms of $N_2O$; this results in the blue shift in the vibrational frequency [2].

For $N$ = 3 to 5, although the increase in $I$ with $N$ is basically from the added contribution of each rigidly attached $^4$He atom, it is found that one needs to consider an increase in the $^4$He-M distance (D) with $N$, as well; Fig. 1(e) shows the required $D$ in the rotor (calculated by using a computer program [45]) which reproduces the experimental value of $B$ for different $N$ (3 to 5). However, increase in $D$, as shown in the graph, is consistent with the fact that strength of $^4$He-$N_2O$ interaction decreases with increasing number of $^4$He atoms attached to the molecule.

Now, the blue shift in $\nu_1$ vibrational frequency continues to be there till $N$ = 5, with a red shift from there on, although, $B$ continues to decrease till $N$ = 6. This clearly indicates that the 6th $^4$He atom is not a part of this equatorial ring, even though it is rigidly attached to the molecule. Consequently, we consider that the 6th $^4$He atom, although forming a part of the rotor, goes to a plane perpendicular to the plane of the equatorial ring (as shown in Fig. 2(a). One may note that this observation is very much in consistence with the fact that each interacting particle ($^4$He atom) should exclusively occupy a space for itself (which is equal to half of its de Broglie wavelength) and thus, $^4$He atoms can not, endlessly, continue to occupy the equatorial ring.

In this context, let $r$ be bond length for $^4$He-$N_2O$. Therefore, the equatorial ring in which $^4$He atoms are accommodated are of length $2\pi r$ (= 6.284$r$). Let the effective size of each $^4$He atom (*i.e.* half of its de Broglie wavelength) be $a$. Thus, the total number of $^4$He atoms that can be accommodated in the ring is 6.284$r/a$. However, in order to determine $r$ and $a$ one must exactly know/ determine the interaction potential among all the particles of the system, which is beyond the scope of the present paper. Thus, without going into any rigorous quantitative explanation for the presence of a maximum of five atoms in the equatorial ring, the corresponding experimental spectroscopic data are considered sufficient to conclude that only five $^4$He atoms are there in the equatorial ring (as discussed above).

Now, as soon as the 7th $^4$He atom comes in the picture, one finds that $B$ does not increase any more and remains almost the same as that for $N$ = 6, but, the shift in the vibrational frequency continues to be red. We, therefore, propose that the 7th $^4$He atom also goes to a position, similar to that of the 6th $^4$He atom. However, we consider that the 6th and 7th $^4$He atoms, together with one of the $^4$He atoms of the equatorial ring, form the said equipotential ring, as shown in Fig. 2(b) (i and ii). Obviously, the ring does not exactly rotate with the rotor ($^4$He$_4$-N$_2$O); however, one may note that as the number of $^4$He atoms in the equipotential ring is small, the ring can not be considered equipotential to a good approximation and thus, it is expected to have some interaction with the rotor, resulting in its $I$ being as high as that of $N = 6$, although the rotor is containing only four $^4$He atoms. In other words, apart from the structure of the rotor, the interaction of the equipotential ring with the rotor also plays an important role for $N = 7$. The same is true for $N = 8$ and $N = 9$, as the corresponding $I$ is almost the same as that of $N = 6$, although, the shift in the vibrational frequency remains red; the rotor still remains $^4$He$_4$-N$_2$O, with extra $_4$He atoms going to the equipotential ring (the structure being shown in Fig. 2(c) and Fig. 2(d), respectively).

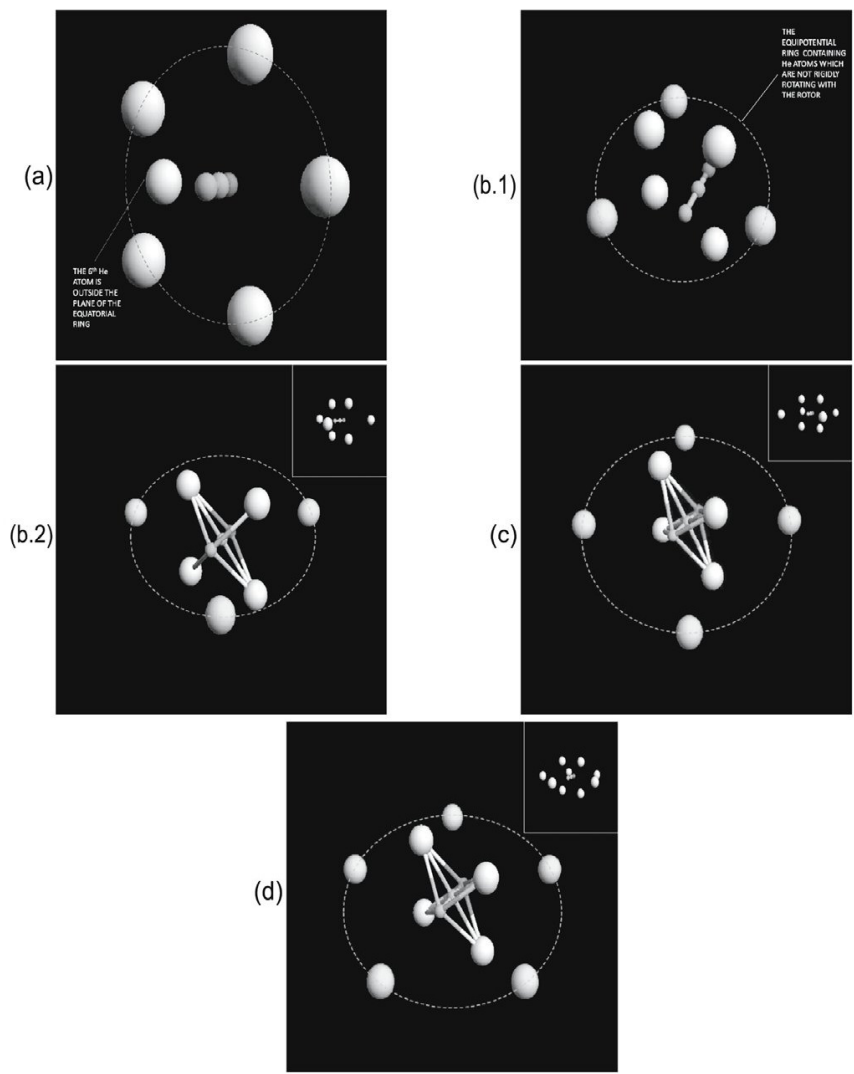

**Figure 2.** (a) 6 $^4$He atoms attached to N$_2$O and rotating with it (6 $^4$He atoms being in the equatorial plane and the 6th $^4$He atom remaining away from it). (b.1) 6th and 7th $^4$He atoms, together with one of the $^4$He atoms of the equatorial ring, forming the said equipotential ring. (b.2) A more symmetric structure for $N = 7$ showing 4 $^4$He atoms (in the equatorial plane) attached to N$_2$O and rotating with it, and 3 $^4$He atoms forming the equipotential ring. (c) Structure for $N = 8$, showing 4 $^4$He atoms (in the equatorial plane) attached to N$_2$O and rotating with it and 4 $^4$He atoms forming the equipotential ring. (d) Structure for $N = 9$, showing 4 $^4$He atom (in the equatorial plane) attached to N$_2$O and rotating with it and 5 $^4$He atoms forming the equipotential ring.

For $N = 10$ and 11, $B$ shows a sudden increase, even though the shift in the vibrational frequency still remains red. So, we propose that 10th and 11th $^4$He atoms go to the equipotential ring. However, as the size of the ring increases due to these added $^4$He atoms, the ring gets very much detouched from the rotor ($^4$He$_4$-N$_2$O), i.e., it does not interact much with the rotor, resulting in a sudden increase in $I$.

The contribution to the decrease in $I$ also comes from the fact that as the number of $^4$He atoms increases in the equipotential ring, the order of the rotational symmetry ($n$) of the potential, $V_n = V_0/2 [1 - \cos(n\phi)]$, also increases which tends to decrease $I$; $n$ being the number of $^4$He atoms in the equipotential ring providing the periodic potential $V_n$ there. This fact can be derived from the analysis of the Mathieu's equation [44], as discussed below.

As discussed in [44], when the height of the potential is reasonably small,

$$S = 7.692(b^{1/2} - 1.0) \tag{1}$$

where $S$ and $b$ are Mathieu's parameters with $S = 4V_0/B_0n^2$ and $b = 4\Delta E/B_0 n^2$; $B_0$ is the rotational constant of the rotor in the absence of the said potential and $\Delta E$ is the difference in energy level between two rotational states when the rotor is rotating in the presence of $V_n$. Thus, one can replace $\Delta E$ by $2B$. Substituting the above values in Eq. (1) one can easily find that

$$B = \frac{2V_0^2}{59.17 n^2 B_0} - \frac{V_0}{7.692} + \frac{n^2 B_0}{8} \tag{2}$$

which clearly shows that as $n$ increases (the number of $^4$He atoms in the equipotential ring), $B$ also increases. Further, increase in $V_0$ also decreases $I$.

For $N = 12$ to 16, one observes that $B$ consistently decreases with increase in $N$, shift in the vibrational frequency remaining red. This can be easily explained, as the extra $^4$He atoms coming in the equipotential ring, obviously, influence the rotor by modifying the $^4$He-N$_2$O bond length and/or its planar structure, consequently, decreasing $B$. In this context, one may note that the effect of the said modification has to be greater than that of the increase in the order of the rotational symmetry (discussed in the last paragraph), resulting in an overall increase in *I*.

Now, with the help of the said computer program [45], we have calculated *D* in the rotor (which produces the experimental value of *B*) for zero angular deformation in the rotor and show the result (for *N* = 12 to 16) in Fig. 3(a). Again, with the help of the computer program in [46], change in the angular position ( ) of the $^4$He atoms in the rotor, for a particular value of $D$, is calculated; the results are shown in Fig. 3(b) (1, 2 and 3). Here,   means the angle by which $^4$He atoms of the equatorial ring are shifted away (along ±y direction) from its planer structure (z-x plane), which produces the experimental value of $B$. While, the range of possible $D$ and   are within the expected range, the systematic increase in $D$ or   of the rotor (for $N$ = 12 to 16) seems to arise due to the interaction of the $^4$He atoms in the equipotential ring with those of the rotor, as argued above.

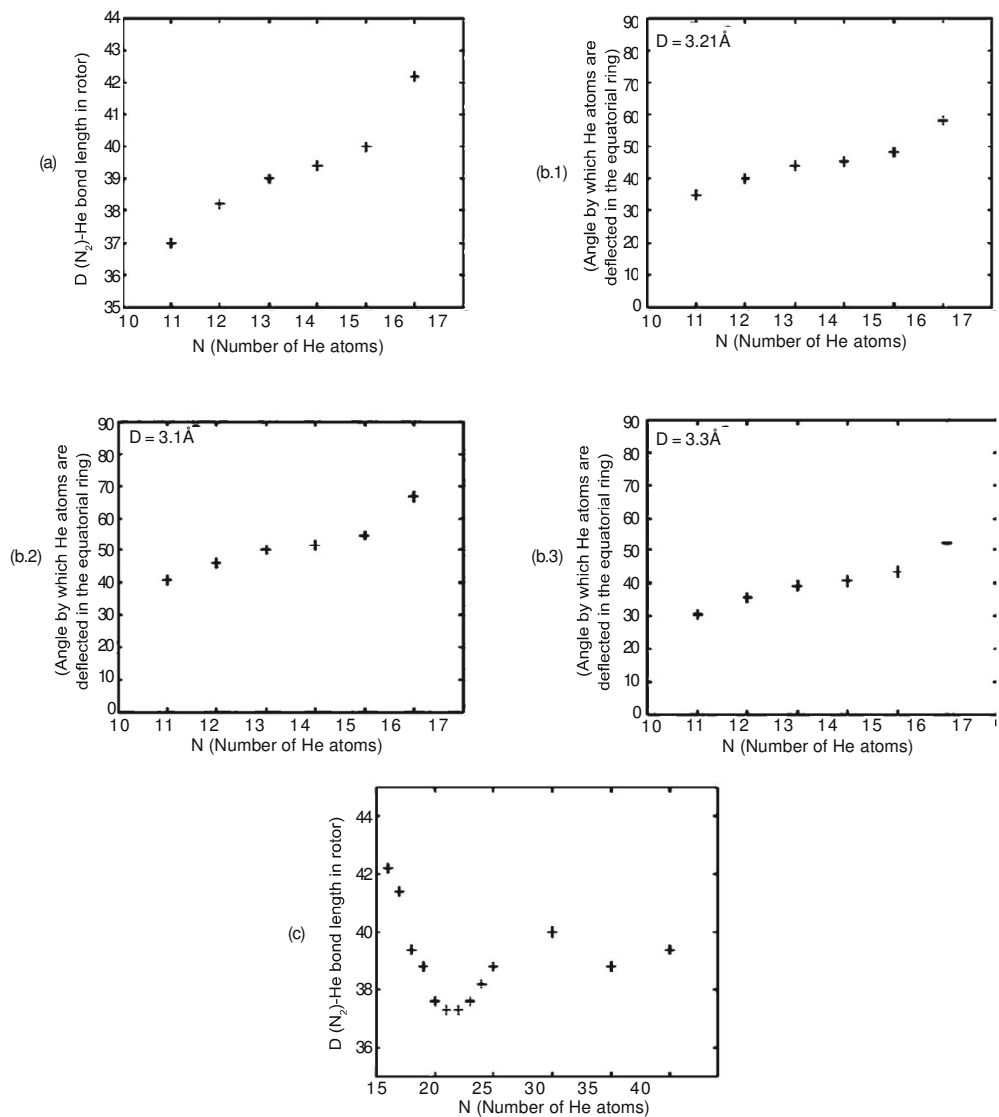

**Figure 3.** (a) Change of D (in Å) with *N* for *N* = 12 − 16. Fig. 3(b.1) Change of (in degree) with *N* for *N* = 12 − 16 (for *D* = 3.21 Å, the value of D calculated in the absence of equipotential ring). Fig. 3(b.2) Change of (in degree) with *N* for *N* = 12 − 16 (for *D* = 3.1 Å) and Fig. 3(b.3) Change of (in degree) with *N* for *N* = 12 − 16 (for *D* = 3.3 Å).

Beyond $N$ = 16, $B$ has a kind of oscillatory change with $N$. Similarly, one may also observe that there is a sharp change in the slope for shift in the vibrational frequency around this point. Both these observations suggest that the equipotential ring gets saturated with about 12 (16-4) $^4$He atoms. It may be noted that the equipotential ring can accommodate more $^4$He atoms than the equatorial ring, because, it surrounds the length of the molecule and not the molecular axis. When $^4$He atoms have occupied all the possible sites (apparently, about 16) close to these two rings (*i.e.*, equatorial ring and equipotential ring), one may visualise the formation of a 3-D saturated shell by these atoms. Any further addition of $^4$He atoms makes a beginning of second shell which will also gets saturated for a certain number of $^4$He atoms leading to the beginning of third shell and so on. In this context, it may be mentioned that a shell can contain only a particular number of $^4$He atoms, because, each $^4$He atom occupies an exclusive volume for itself, as mentioned above. Since the evolution of these shells can make a periodic change in the structure of the rotor with change in $N$, we rightly observe the oscillation of $B$ [2] and corresponding indicators in the shift of 1 frequency [2]. This periodic change in the structure of the rotor, say, its $^4$He-N$_2$O bond length, can be calculated with the help of the computer programs, like that in [45]. Value of $D$ so obtained (for N > 16) is reported in Fig. 3(c).

It may be noted that we have not calculated the required $D$ (and/or  ) for $N$ = 7 to 11, because, the basic reason for the variance among the corresponding values of $B$ for those $N$ is argued to be due to the interaction of the rotor with the equipotential ring (resulting in the hindrance in the free rotation) which is not rigorously considered at a quantitative level. Furthermore, for calculating $D$ (and/or  ) for $N$ = 12 to 16, we have not considered the said hindrance, as well; the values of required $D$ would have been lesser, had we considered the hindrance also, as the later would have accounted for the $I$, to some extent. However, the fluctuation in $D$ so obtained is expected to be similar to that of the true situation (although, each calculated $D$ is having a higher value than that would have been, had we considered the hindrance), for, the hindrance can, approximately, be considered to be constant for higher $N$. Certainly, for $N$ = 3 to 5, the required values of $D$ are the most accurate ones, as we have considered the contribution of all the $^4$He atoms present in the structure and there is no interaction from the equipotential ring ($D$ for $N$ = 6 is not calculated, as, there can be any trade off between the bond length between the 6th $^4$He atom & the N$_2$O molecule and the radius of the equatorial ring).

It is also worth noting that the equipotential ring is a hypothetical ring. Actually, there is a shell of $^4$He atoms surrounding the molecule. The $^4$He atoms that happen to fall in the plane of rotation, render a sort of equipotential ring for the rest of the structure, *i.e.*, rotor . Such a cluster/ shells of $^4$He atoms has been concluded by experiments, *e.g.,* [5], although, the detection of the hypothetical equipotential ring is apparently difficult.

*5.1.1. A slightly alternative approach:*

Instead of the above picture, one may alternatively consider a more quantum mechanical picture of a density cloud of $^4$He atoms, similar to an electron cloud (although, that would also basically give us a similar understanding of the dynamics, as discussed below). For that purpose one needs to calculate the potential energy surface for each added $^4$He atom which has to be, undoubtedly, cylindrically symmetric. So, when only one $^4$He atom is attached to the N$_2$O molecule, one may consider a ring of the cloud of that $^4$He atom around that molecule, as shown in Fig. 4(a). It may be noted that as both an electron and a $^4$He atom at low temperature are quantum particles, one can always talk in terms of probability density rather than a classical particle. However, the electrons in an orbital is slightly different form that of $^4$He atoms in a shell/ ring. An electron, being a Fermion follow Paulis exclusion principle and thus, not more than two electrons can be there in an orbital. $^4$He atoms are Bosons and many atoms can be there in a shell/ ring, although, there number is also limited by the fact that each $^4$He atom, being an interacting particle, occupies an exclusive volume for itself, equal to half of its de Broglie wavelength. Now, as discussed above (what appears from [2]), as successive $^4$He atoms are considered, not more than 5 $^4$He atoms can contribute to the equatorial ring of the $^4$He cloud (Fig. 4(b)), as each interacting $^4$He atom should exclusively occupy a space for itself (which is equal to half of its *de Brogle* wavelength) and thus, $^4$He atoms can not, endlessly, continue to occupy the ring. Now, when the structure rotates about an axis perpendicular to the molecular axis, the ring of $^4$He cloud rotates rigidly with the molecule and contributes to the *I*. With the 6th $^4$He atom in the picture, one can consider it to form a separate ring, towards an end of the molecule, in consistence with [2] (as considered in our earlier argument); all the $^4$He atoms contribute to *I* and the structure is shown in Fig. 4(c).

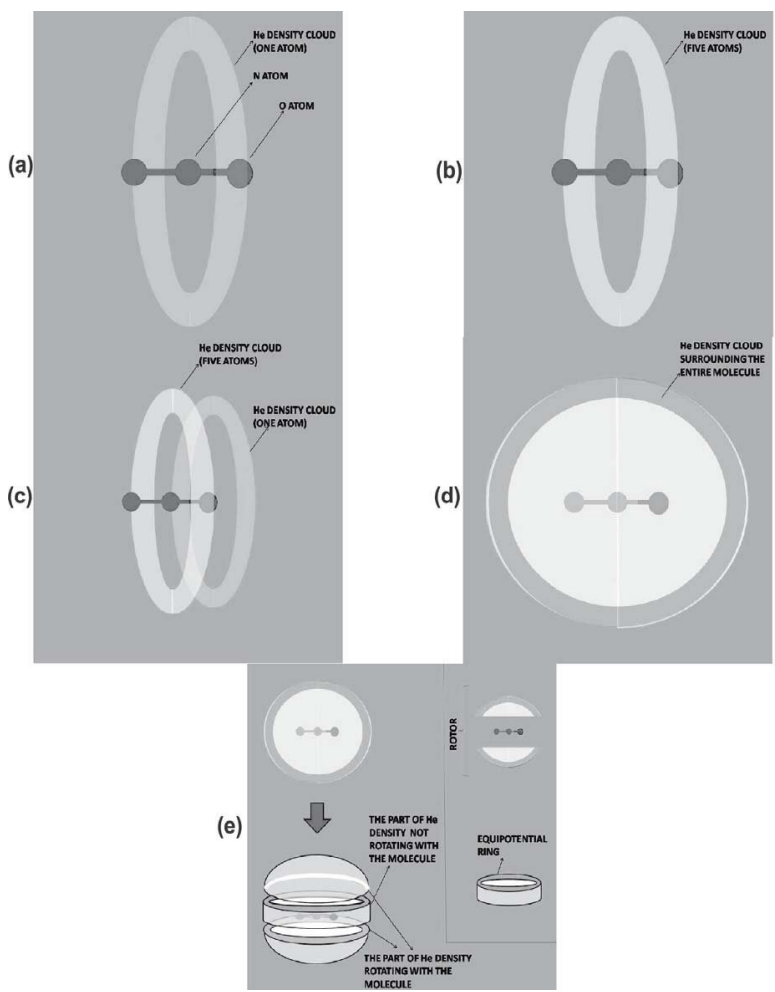

**Figure 4.** (a) A ring of the cloud of a $^4$He atom around a N$_2$O molecule. (b) A ring of the cloud of 5 $^4$He atoms around a N$_2$O molecule. (c) Two rings of the cloud of $^4$He atoms around a N$_2$O molecule (5 $^4$He atoms forming one ring and one $^4$He atom forming the other). (d) 3-D shell of cloud of $^4$He atoms around the N$_2$O molecule. (e) The equipotential ring not taking part in the rotation and only the two hemispheres on either sides of the ring rotating with the molecule.

However, as the 7th $^4$He atom comes, it can be considered to form another ring of $^4$He cloud towards another end; now with this three rings in the picture one can, actually, consider a 3-D shell of cloud of $^4$He atoms around the N$_2$O molecule (as shown in Fig. 4(d). It may be noted the thickness of the shell is, obviously, non uniform along y-axis (although in the figure it is shown to be uniform), which, certainly, tends to be uniform with increase in *N*. Now, in the said shell, when the molecule rotates, it finds a, sort of, uniform density of $^4$He atoms in the plane of its rotation which acts as the said equipotential ring, not taking part in the rotation (similar to what has been discussed earlier) and only the two hemispheres on either sides of the ring rotates with the molecule (Fig. 4(e)). Consequently, the $^4$He atoms in the equipotential ring does not directly contribute to the *I*. However, the interaction of equipotential ring with the rotor (the two hemispheres together with the molecule) has a role in the determination of the magnitude of *I*. It is very important to note that although, here we are talking in terms of density cloud, both the rotor and the equipotential ring must have an integral number of atoms and a value of *I*, corresponding to a particular *N*, is only because of those integral number of atoms in the rotor (having definite bond lengths and bond angles, determining its structure) and its interaction with the equipotential ring.

Thus, we can see that in either of the approaches we have, basically, explained the experimental observations in a similar manner, with the concept of equipotential ring, rotor, *etc.* being the same. The only significant difference between the two approaches is that unlike in the case of later approach, former approach has not considered a density cloud of $^4$He atoms, but classical balls of $^4$He atoms.

*5.2 Analysis of $^4He_N$-HCCCN :*

Identifying the HCCCN molecular axis with the *y* axis and that the molecule rotates about z axis, $^4$He atoms are first considered to occupy positions on a ring (equatorial ring) in x-z plane. These atoms are assumed to have reasonably good binding with the molecule and consequently, the resulting complex rotates like a rigid rotor, as shown in Fig. 5(b), 5(c), 5(d), 5(e), 5(f) and 5(g) for *N* = 3, 4, 5, 6, 7 and 8, respectively. Obviously, *I* of the rotor increases (*B* decreases) with increase in *N* of such atoms, in consistence with [9], approximately.

For *N* = 3 to 8, although the increase in *I* with *N* is basically from the added contribution of each rigidly attached $^4$He atom, it is found that one needs to consider a change in the $^4$He-M distance (*D*) with *N*, as well; Fig. 5(h) and 5(i) show the required *D* in the rotor (calculated by using a computer program, given in [45]) which reproduces the experimental value of rotational transition frequency for *J* = 1 − 0 and *J* = 2 − 1, respectively, for different *N* (3 to 8).

Now, in the case of $^4$He-N$_2$O clusters, we observed that the blue shift in $\nu_1$ vibrational frequency continues to be there till *N* = 5, with a red shift from there on, although, *B* continues to decrease till *N* = 6, indicating that the 6th $^4$He atom is not a part of this equatorial ring, even though it is rigidly attached to the molecule. Consequently, we considered that the 6th $^4$He atom, although forming a part of the rotor, goes to a plane perpendicular to the plane of the equatorial ring. However, in case of HCCCN as the doped molecule, we do not have data for vibrational frequency shift. Thus, we assume that the 9th $^4$He atom in $^4$He$_6$-HCCCN cluster, like the 6th $^4$He atom in $^4$He$_9$-N$_2$O cluster, goes to a plane perpendicular to the plane of the equatorial ring (although, forming a part of the rotor and increasing the rotational transitions frequency, in consistence with [9]); the structure is shown in Fig. 6(a). This is because, $^4$He atoms cannot, endlessly, continue to occupy the equatorial ring, as discussed above.

Now, as soon as the 10th $^4$He atom comes in the picture, one finds that the rotational transition frequency does not increase any more and remains almost the same as that for *N* = 9. We propose that the 10th $^4$He atom also goes to a position, similar to that of the 9th $^4$He atom, considering that the 9th and 10th $^4$He atoms, together with two of the $^4$He atoms of the equatorial ring, form the said equipotential ring, as shown in Fig. 6(b.i) and Fig. 6 (b.ii). Obviously, the ring does not exactly rotate with the rotor ($^4$He$_6$-HCCCN), although, one may note that as the number of $^4$He atoms in the equipotential ring is small, the ring cannot be considered equipotential to a good approximation and thus, it is expected to have some interaction with the rotor. This results in its rotational transition frequency being as high as that of $N = 9$, although, the the rotor is containing only six $^4$He atoms; *i.e.*, apart from the structure of the rotor, the interaction of the equipotential ring with the rotor also plays an important role for $N = 10$. The same is true for $N = 11$, as the corresponding rotational transition frequency is almost the same as that of $N = 9$; the rotor still remains $^4$He$_6$-HCCCN, with extra $^4$He atom going to the equipotential ring (the structure being shown in Fig 6.(c).

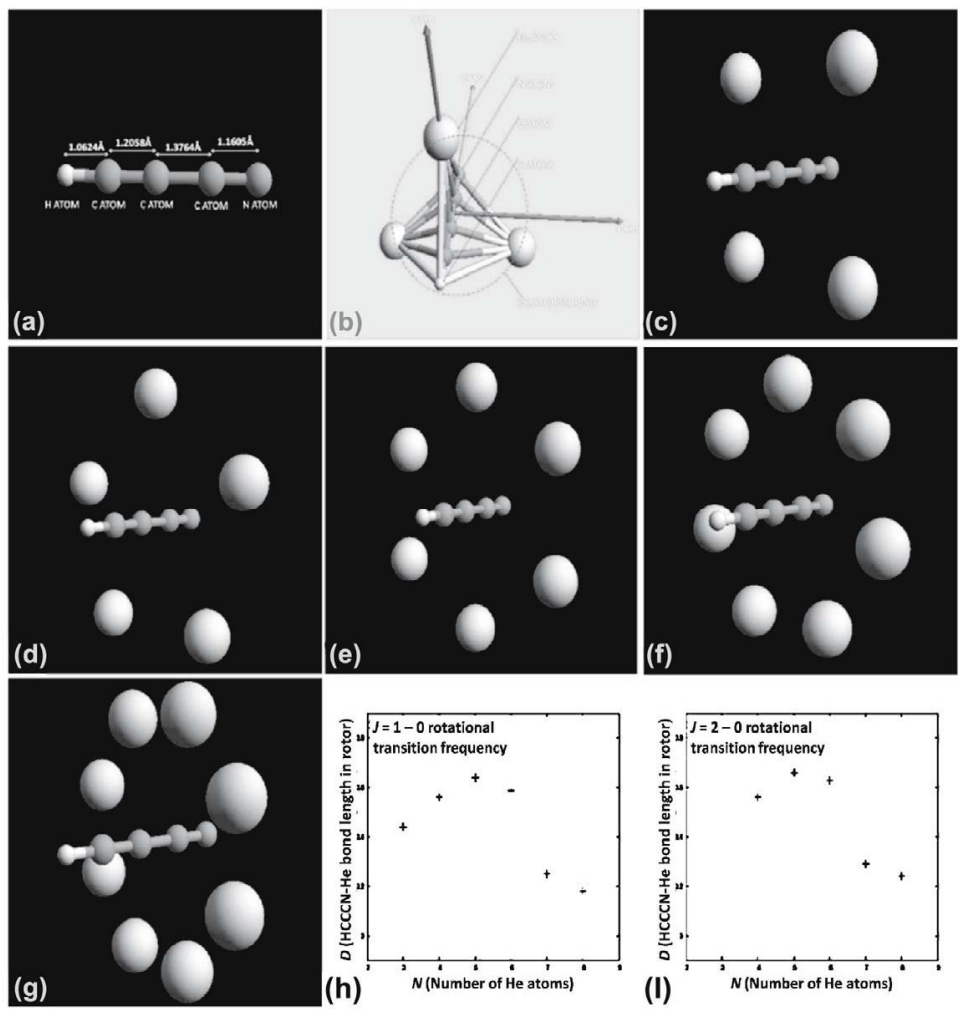

**Figure 5.** (a) The structure of HCCCN molecule. (b) 3 $^4$He atoms [in the equatorial plane (z-x plane)] attached to HCCCN and rotating with it along an axis perpendicular to y axis. (c) 4 $^4$He atoms (in the equatorial plane) attached to HCCCN and rotating with it. (d) 5 $^4$He atoms (in the equatorial plane) attached to HCCCN and rotating with it. (e) 6 $^4$He atoms (in the equatorial plane) attached to HCCCN and rotating with it. (f) 7 $^4$He atoms (in the equatorial plane) attached to HCCCN and rotating with it. (g) 8 $^4$He atoms (in the equatorial plane) attached to HCCCN and rotating with it. (h) Change of D (in Å) with $N$ for $N = 3 - 8$ (for $J = 1 - 0$). (i) Change of D (in Å) with $N$ for $N = 4 - 8$ (for $J = 2 - 1$).

For $N$ = 12 and 13, the rational transition frequencies show a sudden increase. We propose that 12th and 13th $^4$He atoms also go to the equipotential ring, but, as the size of the ring increases due to these added $^4$He atoms, the ring gets very much detouched from the rotor ($^4$He$_6$-HCCCN), *i.e.*, it does not interact much with the rotor, resulting in a decrease in $I$.

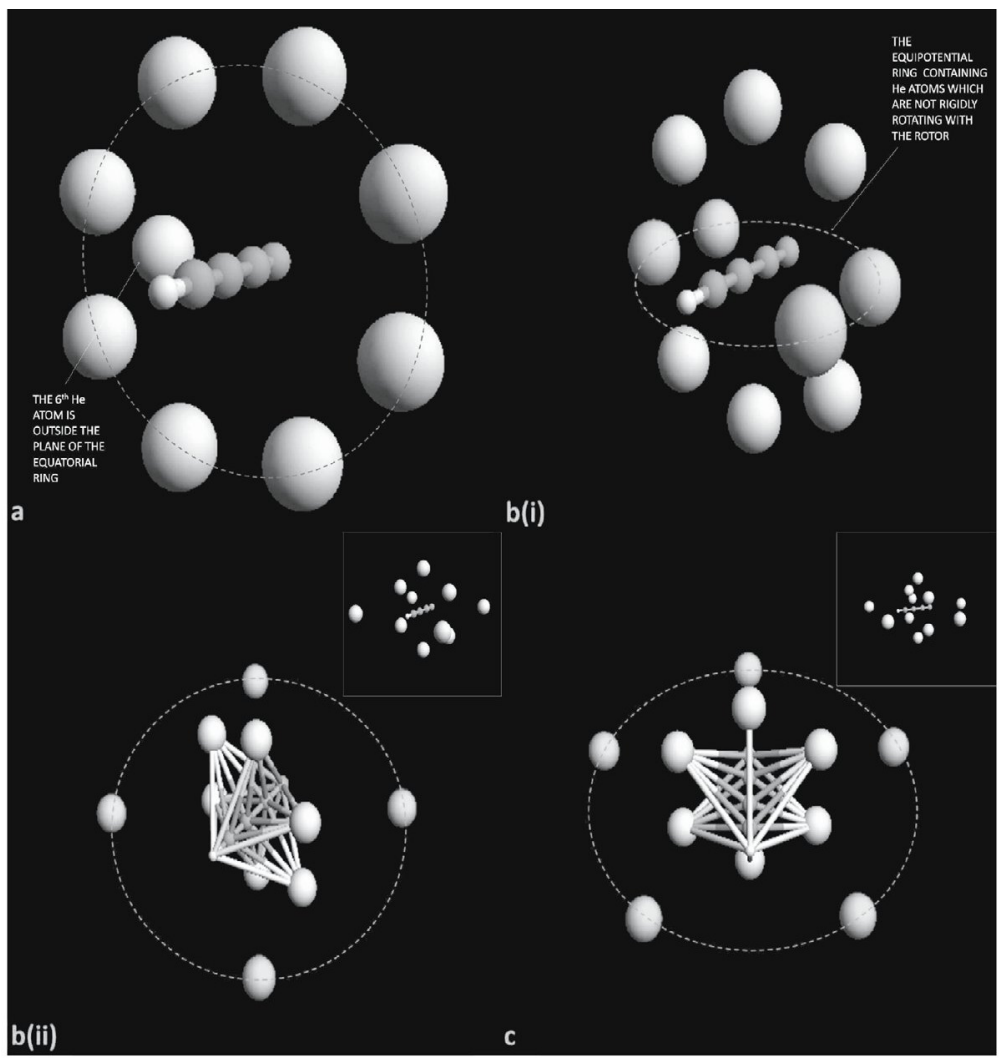

**Figure 6.** (a) 9 $^4$He atoms attached to HCCCN and rotating with it (8 $^4$He atoms being in the equatorial plane and the 9th $^4$He atom remaining away from it). (b(i)) 9th and 10th $^4$He atoms, together with two of the $^4$He atoms of the equatorial ring, forming the said equipotential ring. (b(ii)) A more symmetric structure for $N$ = 10 showing 6 $^4$He atoms (in the equatorial plane) attached to HCCCN and rotating with it, and 4 $^4$He atoms forming the equipotential ring. (c) Structure for $N$ = 11 showing 6 $^4$He atoms (in the equatorial plane) attached to HCCCN and rotating with it and 5 $^4$He atoms forming the equipotential ring.

The contribution to the decrease in *I* also comes from the fact that as the number of $^4$He atoms increases in the equipotential ring, the order of the rotational symmetry ($n$) of the potential, $V_n = V_0/2 [1 - \cos(n\phi)]$, also increases which tends to decrease *I*; we have assumed $V_0 (<< 2B)$. Further, increase in $V_0$ also decreases *I*. This fact is discussed in detail in case of $^4$HeN-N$_2$O (c.f., Eq. 2), with the help of the findings in [44].

For *N* = 14 to 18 [or, more (as, the graph is discontinuous in the region *N* = 19 to 24, as shown in [9])], one observes that the rotational transition frequency consistently decreases with increase in *N*. This is explained by considering that the extra $^4$He atoms, coming in the equipotential ring, obviously, influence the rotor by modifying the $^4$He-N$_2$O bond length and/or its planar structure, consequently, decreasing B. It may be noted that the effect of the said modification has to be greater than that of the increase in the order of the rotational symmetry (discussed in the last paragraph), resulting in an overall increase in *I*.

Now, with the help of the computer program in [45], we calculate *D* in the rotor (which produces the experimental value of the rotational transition frequencies) for zero angular deformation in the rotor and show the results (for *N* = 14 to 18) in Fig. 7(a) and Fig. 7(b), for *J* = 1 − 0 and *J* = 2 − 1, respectively. Again, with the help of the computer program in [46], change in the angular position ($\phi$) of the $^4$He atoms in the rotor, for a particular value of *D*, is calculated; the results are shown in Fig. 7(c) (i and ii) and Fig. 7(d) (i and ii), for *J* = 1 − 0 and *J* = 2 − 1, respectively. Again, while the range of possible *D* and $\dot{\phi}$ are within the expected range, the systematic increase in *D* or $\dot{\phi}$ of the rotor (for *N* = 14 to 18) seems to arise due to the interaction of the $^4$He atoms in the equipotential ring with those of the rotor, as argued above.

Beyond *N* = 25 [or, less], rotational transition frequency starts increasing [and this should mark the beginning of a kind of oscillatory change of it with *N*, for further increase in *N* (the experimental value for which is not available till date), in analogy with $^4$HeN-N$_2$O clusters, discussed in previous subsection]. Consequently, one may suggest that the equipotential ring gets saturated with about 19 [or, less] (25 [or, less] -6) $^4$He atoms. It may be noted that the equipotential ring can accommodate more $^4$He atoms than the equatorial ring, because, it surrounds the length of the molecule and not the molecular axis. When $^4$He atoms have occupied all the possible sites close to these two rings (i.e., equatorial ring and equipotential ring), one may visualise the formation of a 3-D saturated shell by these atoms. Any further addition of $^4$He atoms makes a beginning of second shell which will also get saturated for a certain number of $^4$He atoms leading to the beginning of third shell and so on. This is because, each $^4$He atom occupies an exclusive volume for itself, as mentioned above. Since the evolution of these shells can make a periodic change in the structure of the rotor with change in *N*, we can expect to observe the oscillation of rotational transition frequency with *N*, beginning with an increase form *N* = 25 [or, less]; the increase for a small range (*i.e.*, *N* = 25 to 30) is captured in the experiment. This periodic change in the structure of the rotor, say, its $^4$He-HCCCN bond length, can be calculated with the help of the computer programs, like that in [45].

It may be noted that we have not calculated the required $D$ (and/or $\theta$) for $N = 10$ to 13, because, the basic reason for the variance among the corresponding values of rotational transition frequency for those $N$ is argued to be due to the interaction of the rotor with the equipotential ring (resulting in the hindrance in the free rotation) which is not rigorously considered at a quantitative level. Furthermore, for calculating $D$ (and/or   ) for $N = 14$ to 18, we have not considered said hindrance, as well; the values of required $D$ would have been lesser, had we considered the hindrance also, as the later would have accounted for the $I$, to some extent. However, the fluctuation in $D$ so obtained is expected to be similar to that of true situation (although, each calculated $D$ is having a higher value than what would have been observed, had we considered the said hindrance), for, the hindrance can, approximately, be considered to be constant for higher $N$. Certainly, for $N = 3$ to 8, the required values of $D$ are the most accurate ones, as we have considered the contribution of all the $^4$He atoms present in the structure and there is no interacting from the equipotential ring ($D$ for $N = 9$ is not calculated as there can be any trade off between the bond length between the 9th $^4$He atom & the HCCCN molecule and the radius of the equatorial ring).

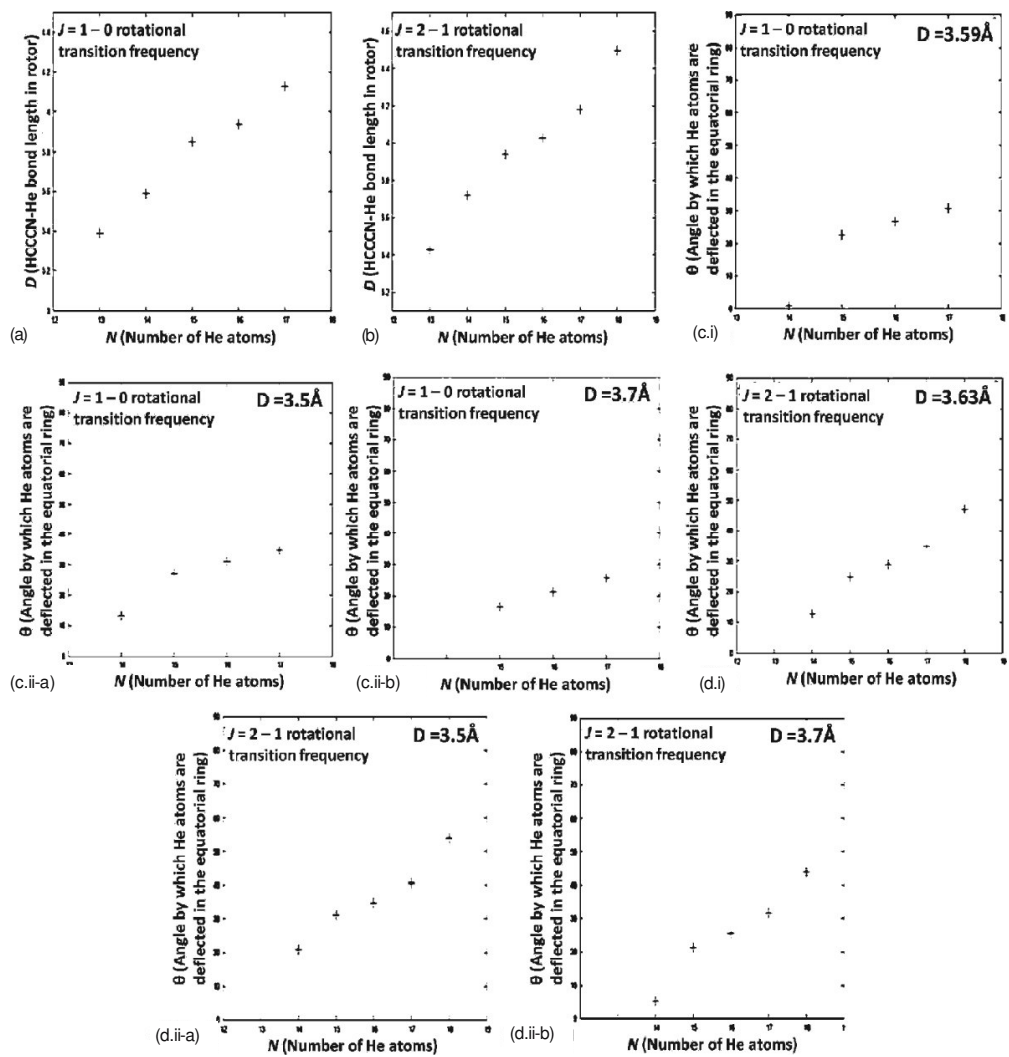

**Figure 7.** (a) Change of $D$ (in Å) with $N$ for $N = 14 - 17$ for $J = 1 - 0$ transition. (b) Change of D (in Å) with $N$ for $N = 14 - 18$ for $J = 2 - 1$ transition. (c(i)) Change of $\theta$ (in degree) with $N$ for N = 14 – 17 for $J = 1 - 0$ transition (for $D = 3.59$ Å, the value of $D$ calculated in the absence of equipotential ring). (c(ii)) Change of (in degree) with $N$ for $N = 14 - 17$, for $J = 1 - 0$ (for $D = 3.5$ Å, 3.7 Å). (d(i)) Change of $\theta$ (in degree) with $N$ for $N = 14 - 18$ for $J = 2 - 1$ transition (for $D = 3.63$ Å, the value of $D$ calculated in the absence of equipotential ring). (d(ii)) Change of (in degree) with $N$ for $N = 14 - 18$, for $J = 2 - 1$ transition (for $D = 3.5$ Å, 3.7 Å).

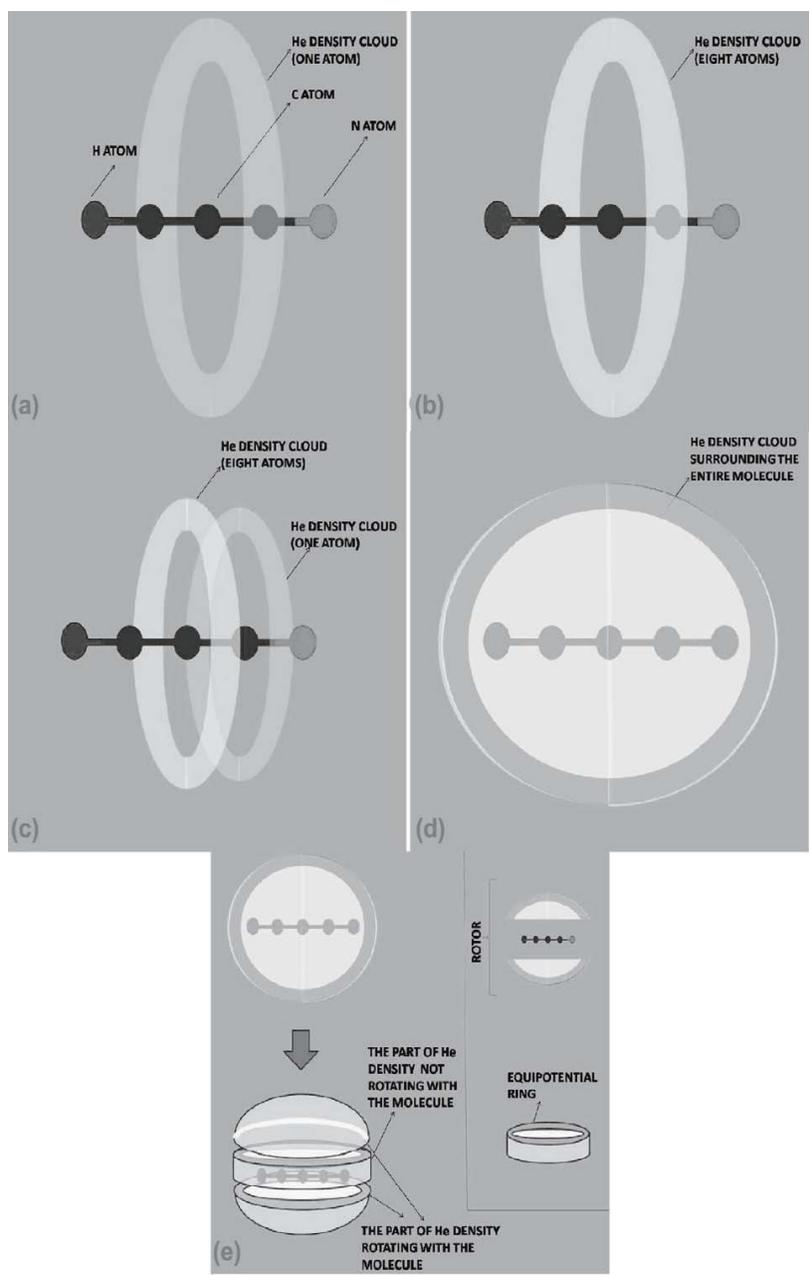

**Figure 8.** (a) A ring of the cloud of a $^4$He atom around a HCCCN molecule. (b) A ring of the cloud of 8 $^4$He atoms around a HCCCN molecule. (c) Two rings of the cloud of $^4$He atoms around a HCCCN molecule (8 $^4$He atoms forming one ring and one $^4$He atom forming the other). (d) 3-D shell of cloud of $^4$He atoms around the HCCCN molecule. (e) The equipotential ring not taking part in the rotation and only the two hemispheres on either sides of the ring rotating with the molecule.

*5.2.1. A slightly alternative approach :*

Here, again, one can use the alternative picture of density cloud of $^4$He atoms (c.f., Fig. 8) and explain the dynamics in a similar manner, as discussed in case of $^4$HeN-N$_2$O.

## 6. Conclusions

N$_2$O is the first molecule where the turn around in change of $B$ with $N$ was first observed. However, till date, no theoretical approach has been able to explain the experimental observations, satisfactorily. However, our approach has explained that the experimental results by sound physical arguments. We have also verified the same with another linear molecule HCCCN.